\documentstyle[12pt,amssymb]{article}

\def\be{\begin{equation}}
\def\ee{\end{equation}}

\normalbaselines
\begin{document}
\begin{center}
\vspace*{1.0cm}

{\LARGE{\bf Quantum Integrable Systems and Special Functions}
\footnote{Published in:\,\,Proc. Clausthal Conference 1999, {\em
Lie Theory and Its Applications in Physics III}, pp.139-154, World
Scientific, Singapore, 2000} }

\vskip 1.5cm

{\large {\bf A.M.Perelomov }
\footnote{\,\,Current e-mail address:
perelomo@dftuz.unizar.es.}}

\vskip 0.5 cm

Institute of Theoretical\\
and Experimental Physics\\
B.Cheremushkinskaya 25 \\
117259 Moscow,  Russia

\end{center}

\vspace{1 cm}

\begin{abstract}\noindent
  The class of quantum integrable systems associated with root systems
was introduced in [OP 1977] as a generalization of the Calogero--Sutherland
systems [Ca 1971], [Su 1972]. For the potential
$v(q)=\kappa (\kappa -1)\sin ^{-2}q$, the wave functions of such systems
 are related to polynomials in $l$ variables ($l$ is a rank of root system)
and they are a generalization of Gegenbauer polynomials and Jack polynomials
[Ja 1970]. In [Pe 1998a],
it was proved that the series for the product of two such polynomials is a
$\kappa $-deformation of the Clebsch--Gordan series. This yields recurrence
relations for these polynomials, in particular, for  generalized zonal
polynomials on symmetric spaces. \\
The present paper follows papers [Pe 1998a], [PRZ 1998]. In last of them,
the recurrence relations were used to compute the explicit expressions
for $A_2$ type polynomials, i.e., for the wave functions of the three-body
Calogero--Sutherland system. \\
As it was shown by  Ragoucy,  Zaugg and the author of this paper
(see [Pe 1999] and Appendix B),
the similar results are also valid in $A_2$ case
for the more general two-parameter deformation ($(q,t)$-deformation)
introduced by Macdonald [Ma 1988].
\end{abstract}

\vspace{1 cm}

%%%%%%
%%%%%%

\section{Introduction}

The class of quantum integrable systems associated with root systems
was introduced in [OP 1977] (see also [OP 1978] and [OP 1983]) as
a generalization of the Calogero--Sutherland systems [Ca 1971], [Su 1972].
Such systems depend on one real parameter $\kappa$ (for root systems of
the type $A_n$, $D_n$ and $E_6$, $E_7$, $E_8$), on two parameters
(for the type $B_n$, $C_n$, $F_4$ and $G_2$) and on three parameters for
the type $BC_n$. These parameters are related to the coupling constants
of the quantum system.

For the potential $v(q)=\kappa (\kappa -1)\,\sin ^{-2}q$ and special values
of  parameter $\kappa$, the wave functions
correspond to the characters of the compact simple Lie groups ($\kappa =1$)
[We 1925/26] or to zonal spherical functions on symmetric spaces
($\kappa =1/2, 2, 4$) [Ha 1958], [He 1978].  At arbitrary values of $\kappa$,
they provide an interpolation between these objects.

This class has many remarkable properties. Here we mention only one:
the wave functions of such systems are a natural generalization of special
functions (hypergeometric functions) to the case of several variables.
The history of this problem and some results may be found in [OP 1983].
In [Pe 1998a], it was shown that the product of two wave functions is a
finite linear combination of analogous functions, namely, of functions
that appear in the corresponding Clebsch--Gordan series. In other words,
this deformation ($\kappa$-deformation) does not change the Clebsch--Gordan
series. For rank 1, we obtain the well-known cases of the Legendre,
Gegenbauer and Jacobi polynomials, and the limiting cases of the Laguerre and
Hermite polynomials (see for example [Vi 1968]). Some other cases were
also considered in [He 1955], [Ja 1970], [Ko 1974], [Ja 1975], [Vr 1976],
[Se 1977], [KS 1978], [Ma 1982], [Pr 1984], [Vr 1984], [HO 1987], [He 1987],
[Op 1988a], [Op 1988b], [Ma 1988], [La 1989], [Op 1989], [St 1989],
[KS 1995] and [CP 1997]. In [PRZ 1998], $\kappa $-deformed
Clebsch--Gordan series was
used  in order to obtain the explicit expressions for the generalized
Gegenbauer polynomials\,\footnote{\,\,In some papers, the name Jack
polynomials [Ja 1970] is used. However, the Jack polynomials are a very
special case of polynomials under consideration. So, we prefer to use the
name generalized Gegenbauer polynomials for the general case and
Jack polynomials for the special case.} of type $A_2$ what gives the
explicit solution of the three-body Calogero--Sutherland model.
For special values of $\kappa =1/2, 2, 4$,
these formulae give the explicit expressions for zonal polynomials
of type $A_2$.

In [Pe 1999] and Appendix B, are presented analogous results
obtained by E. Ragoucy, Ph. Zaugg and the author
for two-parameter family of polynomials of type
$A_2$ introduced by Ruijsenaars [Ru 1987] and Macdonald [Ma 1988].

\section{General description}
\setcounter{equation}{0}

The systems under consideration are described by the Hamiltonian (for more
details, see [OP 1983])
\be
H=\frac{1}{2}\,p^2 + U(q),\qquad p^2=(p,p)=\sum _{j=1}^l\,p_j^2,
\ee
where $p=(p_1,...,p_l),\,\,p_j=-\,i\,\partial /\partial q_j$,
is a momentum operator, and $q=(q_1,...,q_l)$ is a coordinate vector in
the $l$-dimensional vector space $V\sim {\Bbb R}^l$ with standard
scalar product $(\cdot \,,\,\cdot )$.
The potential $U(q)$ is constructed by means of a certain system of vectors
$R^+=\{\alpha\}$ in $V$ (the so-called {\em root systems}):
\be
U=\sum \limits_{\alpha \in R^+}g_\alpha ^2\,v(q_\alpha ),\quad
q_\alpha = (\alpha ,q), \quad g_\alpha ^2=\kappa _\alpha (\kappa _\alpha -1),
\quad g_\alpha =g_\beta ,\quad
\mbox{if}\,\,(\alpha ,\alpha )=(\beta ,\beta ). \ee
Such systems are completely integrable for potentials of five types
(see [OP 1983] for $A_l$; [HO 1987], [He 1987], [Op 1988a], [Op 1988b] and
[Op 1989] for a general case). They are a generalization of the
Calogero--Sutherland systems [Ca 1971], [Su 1972] for which
$\{ \alpha \}=\{e_i-e_j\}$, $\{e_j\}$ being a standard basis in $V$.

In this paper,  we consider in details only the case of $A_2$ with
potential $v(q)=\sin ^{-2}q$.
For the description of other cases see [Pe 1999].

\section{The Clebsch--Gordan series}
\setcounter{equation}{0}

Let us  recall the main results of [Pe 1998a] and specialize them
to the $A_2$ case with potential $v(q)=\sin ^{-2}q$.

The Schr{\"o}dinger equation for this quantum system has the form
\be
H\,\Psi ^\kappa = E(\kappa)\,\Psi ^\kappa ;\qquad
H=-\,\Delta _2+U(q_1,q_2,q_3), \qquad
\Delta _2=\sum \limits_{j=1}^{3}\,\frac{\partial ^2}{\partial q_j^2}
\label{schrod}
\ee
with potential
\be
U(q_1,q_2,q_3)= \kappa(\kappa-1)\left( \sin^{-2}(q_1-q_2)+\sin^{-2}(q_2-q_3)+
\sin^{-2}(q_3-q_1) \right).
\ee
The ground state wave function and its energy are
\be
\Psi _0^\kappa (q)=\left( \prod _{j<k}^3 \,\sin (q_j-q_k)
\right)^\kappa, \qquad E_0(\kappa )= 8\,\kappa^2.
\ee
Substituting $\Psi _\lambda ^\kappa =\Phi _\lambda ^\kappa \,\Psi
_0^\kappa$ in (\ref{schrod}), we obtain
\be
-\Delta ^\kappa \,\Phi _\lambda ^\kappa =\varepsilon _\lambda (\kappa )\,
\Phi _\lambda ^\kappa ,\qquad
\Delta ^\kappa=\Delta _2+\Delta _1^\kappa ,\qquad
\varepsilon _\lambda (\kappa )=E_\lambda (\kappa )-E_0(\kappa ).
\ee
Here the operator $\Delta _1^\kappa $ takes the form
\be
\Delta _1^\kappa =\kappa\sum _{j<k}^3\, \cot (q_j-q_k)
\left( \frac{\partial}{\partial q_j} - \frac{\partial}{\partial
q_k}\right) .
\ee
It is easy to see that the set of symmetric polynomials in variables
$\exp(2iq_j)$ is invariant under the action of $\Delta ^\kappa $.
Such polynomial $m_\lambda$ is labelled by the $SU(3)$ highest weight
$\lambda=m \lambda_1+n \lambda_2$, with $m,n$ being non-negative integers,
and $\lambda_{1,2}$ being two fundamental weights. In general,
\be
\Phi_\lambda^\kappa = \sum_{\mu \leq \lambda}C^\mu_\lambda(\kappa)\,m_\mu,
\quad \mu ,\lambda \in P^+,\qquad m_\mu = \sum_{\nu \in W\cdot \mu}
e^{2i(q,\nu )},
\ee
where $P^+$ denotes the cone of dominant weights, $W$ is the Weyl group,
and $C^\mu_\lambda(\kappa)$ are some constants.

As it was shown in [Pe 1998a], the product of two wave functions is
a finite sum of wave functions (a sort of the $\kappa$-deformed
Clebsch--Gordan series):
\be
\Phi_\mu^\kappa \,\Phi_\lambda^\kappa = \sum_{\nu \in D_\mu(\lambda )}
C_{\mu \lambda }^\nu (\kappa )\,\Phi _\nu ^\kappa.
\label{main} \ee
In this equation, $D_\mu(\lambda)=(D_\mu+\lambda)\cap P^+$, where
$D_\mu$ is a weight diagram of the representation with the highest weight
$\mu$.

Since $\Phi_\mu^\kappa$ are symmetric functions of $\exp(2iq_j)$,
it is convenient to use a new set of variables:
\be
\begin{array}{l}
z_1 = e^{2i q_1}+e^{2i q_2}+e^{2i q_3}, \\
z_2= e^{2i(q_1+q_2)}+e^{2i (q_2+q_3)}+e^{2i (q_3+q_1)}, \\
z_3= e^{2i(q_1+q_2+q_3)}. \end{array} \ee In the centre-of-mass
frame ($\sum_i q_i=0$), the wave functions depend only on two
variables chosen as $z_1$ and $z_2$ (in this case, $z_3=1$). In
these variables, up to a normalization factor, we have \be
\Delta^\kappa =(z_1^2-3z_2)\,\partial _1^2+(z_2^2-3z_1)\,
\partial _2^2+(z_1z_2-9)\,\partial _1\partial _2+(3\kappa +1)\,
(z_1\partial _1+z_2\partial _2),
\ee
where  $\partial _i =\partial/\partial z_i$. Corresponding eigenvalues are
\be
\varepsilon_{m,n}(\kappa) = m^2+n^2+m n+3\,\kappa(m+n).
\ee

We shall use the normalization for polynomials $\Phi_\lambda^\kappa$ such
that the coefficient at the highest monomial is equal to one. Denoting them
by $P^\kappa_{m,n}$, we have
\be
P^\kappa_{m,n}(z_1,z_2) = \sum_{p,q} C^{p,q}_{m,n}(\kappa)\,z_1^p\,z_2^q
= z_1^m\,z_2^n + \mbox{lower~terms},
\label{P-struct}
\ee
with $p+q \geq m+n$ and $p-q \equiv m-n\,(\mbox{mod}~3)$.
As it is easy to see, the first polynomials are
\be
P^\kappa_{0,0}=1, \qquad P^\kappa_{1,0}=z_1, \qquad P^\kappa_{0,1}=z_2.
\ee
Simple consequences of (\ref{main}) for $P^\kappa_\lambda=P^\kappa_{1,0}$
or $P^\kappa_{0,1}$ are [Pe 1998]
\begin{eqnarray}
z_1\,P_{m,n}^\kappa &=& P_{m+1,n}^\kappa +a_{m,n}(\kappa)\,
P_{m,n-1}^\kappa +c_{m}(\kappa) \,P_{m-1,n+1}^\kappa ,
\label{rec-z1} \nonumber \\
&&\\
z_2\,P_{m,n}^\kappa &=& P_{m,n+1}^\kappa +\tilde{a}_{m,n}(\kappa)\,
P_{m-1,n}^\kappa +c_{n}(\kappa) \,P_{m+1,n-1}^\kappa ,\nonumber
\label{rec-z2} \end{eqnarray}
where
\begin{eqnarray}
a_{m,n}(\kappa)&=&\tilde{a}_{n,m}(\kappa) = c_n(\kappa)\,
c_{m+n+\kappa}(\kappa), \nonumber \\
&&\\
c_m(\kappa) &=& \frac{e(m)}{e(\kappa+m)},\qquad e(m)~ =~ \frac{m}{m-1+\kappa}.
\nonumber \end{eqnarray}
Below we shall construct such polynomials using these
recurrence relations.

%section{$A_1$ case}

\section{$A_2$ case}
\setcounter{equation}{0}

Now we proceed to the case of $A_2 \sim su(3)$. In this case,
the representation $d$ of $A_2$ is characterized by two non-negative numbers
$d=d_{mn}$. We have two fundamental representations
\[
d_{10},\qquad d_{01};\qquad (\mbox{dim}\,d_{10}=\mbox{dim}\,d_{01}=3), \]
and also representations
\[
d_{n0},\qquad d_{0n};\qquad \left( \mbox{dim}\,d_{n0} =\mbox{dim}\,d_{0n}=
\frac12\,(n+1)(n+2)\right) ,\]
\[ d_{n1}\,\Big( \mbox{dim}\,d_{n1}=(n+1)(n+3)\Big) ,\qquad d_{nn}\,
\Big( \mbox{dim}\,d_{nn}=(n+1)^3\Big) .\]

We start with the Clebsch--Gordan series
\be \begin{array}{l}
d_{10}\otimes d_{n+1,0}=d_{n+2,0}\oplus d_{n,1},\\
d_{01}\otimes d_{n0} =d_{n,1}\oplus d_{n-1,0}. \end{array} \ee
Excluding $d_{n,1}$, we obtain
\[
d_{n-1,0}\ominus (d_{01}\otimes d_{n0})\oplus (d_{10}\otimes d_{n+1,0})
\ominus d_{n+2,0}=0, \]
or
\be
\chi _{n-1,0}-z_2\,\chi _{n,0}+z_1\,\chi _{n+1,0}-\chi _{n+2,0}=0, \ee
where are introduced the notations:
\be \begin{array}{l}
z_1 =\chi _{10} = e^{i\theta _1}+e^{i\theta _2}+e^{i\theta _3}, \\
z_2= \chi _{01} = e^{-i\theta _1}+e^{-i\theta _2}+e^{-i\theta _3}.
\end{array} \ee
From this, we obtain the expression for the generating function
\begin{eqnarray}
F_0^1(z_1,z_2; u)&=&\sum _{n=0}^\infty \,\chi _{n0}(z_1,z_2)\,u^n,
\nonumber \\
F_0^1(z_1,z_2; u)&=&\left( 1-z_1u+z_2\,u^2-u^3\right) ^{-1}.
\end{eqnarray}

Let us define now the $\kappa $-deformed functions $\tilde P _{n,0}^\kappa
(z_1,z_2)$ by the formula
\be F^\kappa (z_1,z_2;\,u)=\left( 1-z_1\,u+z_2\,u^2-u^3\right) ^{-\kappa }
=\sum _{n=0}^\infty \,\tilde P_{n,0}^\kappa (z_1,z_2)\,u^n.\ee

Differentiating $F^\kappa $ on $u$, $z_1$ and $z_2$, we get
\[ F_u^\kappa =\kappa \left( z_1-2z_2\,u+3u^2\right) F^{\kappa +1}, \]
\[ \begin{array}{ll}
F_{z_1,z_1}^\kappa =\kappa (\kappa +1)u^2\,F^{\kappa +2},\qquad &
F_{z_2,z_2}^\kappa =\kappa (\kappa +1)u^4\,F^{\kappa +2},\\
&\\
F_{z_1,z_2}^\kappa =-\,\kappa (\kappa +1)u^3\,F^{\kappa +2},\qquad &
u\,F_u^\kappa =\kappa \left( z_1u-2\,z_2u^2+3\,u^3\right) F^{\kappa +1},
\end{array} \]
and
\be
\left( 1-z_1u+z_2\,u^2-u^3\right) F_u^\kappa =\kappa \left( z_1-2\,z_2u+
3\,u^2\right) F^\kappa .
\ee
From this it follows the important recurrence formula
\be (n+3)\,\tilde P_{n+3,0}^\kappa =(n+2+\kappa )z_1\,
\tilde P_{n+2,0}^\kappa -(n+1+2\kappa )z_2\,\tilde P_{n+1,0}^\kappa +
(n+3\kappa )\,\tilde P_{n,0}. \ee

We have also
\be F_{z_1}^\kappa =\kappa u\,F^{\kappa +1} ,
F_{z_2}^\kappa =-\,\kappa u^2\,F^{\kappa +1}.\ee

Hence
\be
\partial _{z_1}P_{n,0}^\kappa =n\,P_{n-1,0}^{\kappa +1},
\partial _{z_2}P_{n,0}^\kappa =-\,\frac{n(n-1)}{\kappa +n-1}\,
P_{n-2,0}^{\kappa +1}.
\ee

Finally we have the basic differential equation for $F^\kappa (z_1,z_2;\,u)$:
\begin{eqnarray*}
&&\left( \left( D_{z_1}^2+D_{z_2}^2+D_{z_1}D_{z_2}\right)-3\,z_2
\partial _{z_1}^2-3\,z_1\partial _{z_2}^2-9\,\partial _{z_1}\partial _{z_2}
+3\,\kappa (D_{z_1}+D_{z_2})\right) F^\kappa \\
&=&\left( D_u^2+3\,\kappa D_u\right) F^\kappa (z_1,z_2;\,u); \end{eqnarray*}
\[
D_{z_1}=z_1\,\partial _{z_1},\qquad D_{z_2}=z_2\,\partial z_2,\qquad
D_u=u\,\partial _u.\]

Let us note that the normalization of polynomials $\tilde P_{n,0}^\kappa
(z_1,z_2)$ follows from the expression (4.5) for the generating function.
Namely,
\be
\tilde P_{n0}^\kappa (z_1,z_2)=\frac{(\kappa )_n}{n!}\,z_1^n+\cdots =
\frac{(\kappa )_n}{n!}\,P_{n,0}^\kappa (z_1,z_2), \ee
where
\[ (\kappa )_n=(\kappa )(\kappa +1)\cdots (\kappa +n-1).\]

The main property of this normalization is that $\tilde P_{n,0}^\kappa
(z_1,z_2)$ has a polynomial dependence on the parameter $\kappa $.

Now we shall consider other Clebsch--Gordan series for $\kappa =1$:
\[ d_{1,0}\otimes d_{n+1,0}=d_{n+2,0}\oplus d_{n,1}.\]
According to [Pe 1998], the analogous formula is valid for an arbitrary value
of $\kappa $, i.e.,
\be
a_nz_1\,\tilde P_{n+1,0}^\kappa =b_n\,\tilde P_{n+2,0}^\kappa +c_n\,
\tilde P_{n,1}^\kappa . \ee
Here
\[
a_n=\kappa +n+1,\qquad b_n=n+2, \]
and we have
\be
c_n(\kappa )\,\tilde P_{n,1}^\kappa =(\kappa +n+1)z_1\,\tilde P_{n+1,0}^
\kappa -(n+2)\,\tilde P_{n+2,0}^\kappa .\ee

Let us calculate now the generating function for both left-hand and
right-hand sides of this equation,
\begin{eqnarray}
G^\kappa &=&\sum _{n=0}^\infty \,c_n(\kappa )\,\tilde P_{n,1}^\kappa
(z_1,z_2),\nonumber \\
&&\\
G^\kappa &=&\frac{\kappa z_1}u\,(F_0^\kappa -1)+\frac{z_1}u\,F_1^\kappa
-\frac1{u^2}\,(F_1^\kappa -\kappa z_1u),\nonumber \end{eqnarray}
where
\[
F_0^\kappa =\sum _{n=0}^\infty \tilde P_{n,0}^\kappa \,u^n;\qquad F_1^\kappa =
\sum _{n=0}^\infty n\,\tilde P_{n,0}^\kappa \,u^n=D_u\,F_0^\kappa , \]
and
\[
G^\kappa = \frac1{u^2}\,(\kappa z_1u-(1-z_1u)D_u)\,F_0^\kappa . \]
Finally,
\be
G^\kappa =\kappa \left( 2z_2-(z_1z_2+3)\,u+2\,z_1u^2\right) F_0^{\kappa +1}.
\ee
From this, it follows three-term recurrence relation
\[
\tilde P_{n,1}^\kappa =\kappa \left( 2\,z_2\,\tilde P_{n,0}^{\kappa +1}-
(z_1z_2+3)\,\tilde P_{n-1,0}^{\kappa +1}+2\,z_1\,\tilde P_{n-2,0}
^{\kappa +1}\right) . \]

Now let us follow [PRZ 1998] and construct the general polynomials in terms
of the simplest polynomials (Jack polynomials) $P^\kappa_{m,0}$ and
$P_{0,n}^\kappa $. We get
\be
P^\kappa_{m,0}\,P^\kappa_{0,n} = \sum_{i=0}^{{\rm min}(m,n)}
\gamma^i_{m,n}\,P^\kappa_{m-i,n-i}.
\label{ppgammap} \ee
This is a consequence of equation (\ref{main}), with the notable difference
that the sum on the right-hand side is over a restricted domain
(actually, it parallels exactly the $SU(3)$ Clebsch--Gordan decomposition).

To prove this, let us assume that (\ref{ppgammap}) is valid up to
$(m,n)$.  Then using (\ref{rec-z1}) and (\ref{rec-z2}), we get
\be
P^\kappa_{m,0}\,P^\kappa_{0,n+1} = \sum_{i=0}^{{\rm min}(m,n+1)}
\gamma^i_{m,n+1}\,P^\kappa_{m-i,n+1-i} + c_n\,\delta^i_{m,n+1}\,
P^\kappa_{m+1-i,n-1-i},
\label{ppgammadeltap} \ee
where we defined
\begin{eqnarray}
\gamma^i_{m,n+1} &=& \gamma^i_{m,n} + \tilde{a}_{m-i+1,n-i+1} \,
\gamma^{i-1}_{m,n} - c_n \, c_{m-i+1} \, \gamma^{i-1}_{m,n-1},
\label{gammadelta1} \\
\delta^i_{m,n+1} &=& c_n^{-1} \, c_{n-i} \, \gamma^i_{m,n} -
\gamma^i_{m,n-1} - a_{m-i+1,n-i} \, \gamma^{i-1}_{m,n-1} \nonumber \\
& & \mbox{} + c_{n-1} \, c_{\kappa+n-1} \, \gamma^{i-1}_{m,n-2}.
\label{gammadelta2} \end{eqnarray} From the polynomial
normalization, we already know that $\gamma^0_{m,n}=1$. After a
straightforward computation, the solution to (\ref{gammadelta1})
proved to be equal \be \gamma^i_{m,n} =
\frac{e(2\kappa+m+n+1-i)_{-i}}{e(1)_i \, e(m+1)_{-i}\,
e(n+1)_{-i}}\,, \label{sol-gamma} \ee where
\[ e(m)=\frac{m}{m-1+\kappa }\,. \]
It implies that $\delta^i_{m,n+1}=0$ in (\ref{gammadelta2}). Let us give
also the more explicit expression\,\footnote{\,\,Note that this expression
for $\gamma _{m,n}^i$ may be obtained from the general Macdonald formula
[Ma 1995], however, the way of proof given in [PRZ 1998] is more convenient
here.}
\be \label{sol-gamma1}
\gamma^i_{m,n} =\frac{(\kappa )^i(m)_i\,(n)_i\,(3\kappa +m+n-1-i)_i}
{i!\,(\kappa +m-1)_i\,(\kappa +n-1)_i\,(2\kappa +m+n-i)_i}\,, \ee
where
\be \begin{array}{l}
(x)^i=x(x+1)\cdots (x+i-1), \\
(x)_i= x(x-1)\cdots (x-i+1). \end{array} \ee

The constructive aspect of this formula is in its inverted form.

\medskip\noindent
{\bf Theorem 1 [PRZ 1998]}. {\em The generalized Gegenbauer polynomials}
$P_{m,n}^\kappa$ {\em of type} $A_2$ {\em are given by the formula}
\be
P^\kappa_{m,n} = \sum_{i=0}^{{\rm min}(m,n)} \beta^i_{m,n}\,
P^\kappa_{m-i,0}\,P^\kappa_{0,n-i}\,,
\label{ppbetap} \ee
{\em where the constants} $\beta _{m,n}^i$ {\em are}
\be
\beta ^i_{m,n}=\frac{(-1)^i}{i!}\,\frac{3\kappa +m+n-2i}{3\kappa +m+n-i}\,
\frac{(m)_i\,(n)_i\,(\kappa )_i\,(3\kappa +m+n-1)_i}{(\kappa +m-1)_i
(\kappa +n-1)_i(2\kappa +m+n-1)_i}\,.\ee

Note that  $\beta^i_{m,n}$ are obtained by using of the relation
\be
\beta _{m,n}^i=-\,\sum_{j=0}^{i-1}\beta^j_{m,n}\,\gamma^{i-j}_{m-j,n-j}\,.
\ee
From this theorem, we see that the construction of a general polynomial
$P^\kappa_{m,n}$ is similar to the construction of $SU(3)$ representations
from tensor products of two fundamental representations.

Likewise, we can consider other types of decompositions,
such as
\be
P^\kappa_{m,0}\,P^\kappa_{n,0} = \sum_{i=0}^{{\rm min}(m,n)} \tilde
\gamma ^i_{m,n}\,P^\kappa_{m+n-2i,i}.
\label{ppgammatp} \ee
The proof is analogous to (\ref{ppgammap}) (see footnote 3).
The coefficients $\tilde \gamma _{m,n}^i$ are given by the formula
\be
\tilde \gamma ^i_{m,n} = \frac{(\kappa )^i\,(m)_i\,(n)_i\,
(2\kappa +m+n-1-i)_i}
{i!\,(\kappa +m-1)_i\,(\kappa +n-1)_i\,(\kappa +m+n-i)_i}. \ee

\medskip\noindent
{\bf Theorem 2 [PRZ 1998]}. {\em There is another formula for polynomials}
$P_{m,n}^\kappa$ {\em at} $m\geq n$:
\be
\tilde\gamma^n_{m+n,n}\,P_{m,n}^\kappa =\sum _{i=0}^n\,\tilde
\beta _{m,n}^i\,P^\kappa_{m+n+i,0}\,P^\kappa_{n-i,0},
\ee
{\em where}
\be
\tilde \beta ^i_{m,n}= \frac{(-1)^i}{i!}\,(\kappa )_i\,\frac{(m+2i)}m\,
\frac{(\kappa +m+n)^i}{(m+n+1)^i}\,\frac{(m)^i}{(\kappa +m+1)^i}\,
\frac{(n)_i}{(\kappa +n-1)_i}\,. \ee

This theorem follows directly from equation (\ref{ppgammatp}). The
coefficients $\tilde\beta^i_{m,n}$ are found by using of the relation
\be
\tilde\beta^i_{m,n} =-\left(\tilde\gamma^{n-i}_{m+n+i,n-i}\right)^{-1}\,
\sum_{j=0}^{i-1} \, \tilde\beta^j_{m,n} \, \tilde \gamma^{n-i}_{m+n+j,n-j} .
\ee

As a by-product, let us specialize equation (\ref{ppbetap}) to the case
$\kappa=1$, where $P^\kappa_{m,n}$ are nothing but the $SU(3)$ characters.
We get
\be
P^{1}_{m,n}= P^{1}_{m,0}\,P^{1}_{0,n}-P^{1}_{m-1,0}\,P^{1}_{0,n-1}.
\ee
From this, we easily deduce the generating function for $SU(3)$
characters (see e.g. [PS 1978])
\be
G^{1}(u,v)=\sum_{m,n=0}^\infty u^m\,v^n\,P^{1}_{m,n} =
\frac{1-uv}{(1-z_1\,u+z_2\,u^2-u^3)\,(1-z_2\,v+z_1\,v^2-v^3)}.
\ee

Closing this section, let us note that for $\kappa $=1/2, 1, 2 and 4,
the obtained formulae yield the explicit expression of zonal polynomials for
certain symmetric spaces listed below.

Let us mention also the papers [CP 1997], [Pe 1998b] where the integral
representations for the case N=3 were obtained.

%% \newpage
\appendix
\section*{Appendix A. List of polynomials $P_{m,n}^\kappa $ with
$m+n\leq 4$}
\setcounter{equation}{0}

Following [PRZ 1998], we list here the polynomials $P_{m,n}^\kappa$
with $m+n \leq 4$:
%\be
\begin{eqnarray}%{lcl}
P_{2,0}^\kappa &=& z_1^2 - \frac{2}{\kappa+1} \,z_2, \nonumber \\
P_{1,1}^\kappa &=& z_1 z_2 -\frac{3}{2\kappa+1}, \nonumber \\
P_{3,0}^\kappa &=& z_1^3 - \frac{6}{\kappa+2} \,z_1 z_2 + \frac{6}
{(\kappa+1)(\kappa+2)}, \nonumber \\
P_{2,1}^\kappa &=& z_1^2 z_2 -\frac{2}{\kappa+1}\, z_2^2 -
\frac{3\kappa+1}{(\kappa+1)^2} \,z_1, \nonumber \\
P_{4,0}^\kappa &=& z_1^4 - \frac{12}{\kappa+3} \,z_1^2 z_2 + \frac{12}
{(\kappa+2)(\kappa+3)}\,z_2^2 + \frac{24}{(\kappa+2)(\kappa+3)}\, z_1,
\nonumber \\
P_{3,1}^\kappa &=& z_1^3 z_2 - \frac{6}{\kappa+2} \,z_1 z_2^2 -
\frac{3(3\kappa+2)}{(\kappa+2)(2\kappa+3)} \,z_1^2
+\frac{30}{(\kappa+2)(2\kappa+3)} \,z_2, \nonumber \\
P_{2,2}^\kappa &=& z_1^2 z_2^2 - \frac{2}{\kappa+1} \,(z_1^3 + z_2^3)
 - \frac{12(\kappa-1)}{(\kappa+1)(2\kappa+3)}\,z_1 z_2 +
 \frac{9(\kappa-1)}{(\kappa+1)^2(2\kappa+3)}\,.\nonumber
\end{eqnarray}

\appendix
\section*{Appendix B. Some formulae for Macdonald polynomials for $A_2$ case
(by A.M. Perelomov, E. Ragoucy and Ph. Zaugg)}
\setcounter{section}{2}
\setcounter{equation}{0}

Here we give only some necessary for us information. For other results and
details, see [Ma 1995].

The Macdonald polynomials of type $A_2$ may be defined as polynomial
eigenfunctions of the Macdonald difference equation
\be
M^1\,P^{(q,t)}_{m,n}(x_1,x_2,x_3)=\lambda \,P^{(q,t)}_{m,n}(x_1,x_2,x_3),
\qquad P_{m,n}=z_1^mz_2^n+\mbox{lower\,\,terms}, \ee
\[ z_1=x_1+x_2+x_3,\qquad z_2=x_1x_2+x_2x_3+x_3x_1,\qquad z_3=x_1x_2x_3,\]
where
\be
M^1 = \prod _{j\neq k}\frac{(tx_j-x_k)}{(x_j-x_k)}\,T_j,\qquad
T_1\,f(x_1,x_2,x_3) = f(qx_1,x_2,x_3), \ldots .\ee

The Macdonald polynomials satisfy the following recurrence relation
\be
z_1\,P^{(q,t)}_{m,n}(z_1,z_2,z_3)=P^{(q,t)}_{m+1,n}+a_{m,n}(q,t)\,P^{(q,t)}
_{m,n-1} + b_{m,n}(q,t)\,P_{m-1,n+1}^{(q,t)}, \ee
where
\begin{eqnarray}
a_{mn} &=& c_n\,\tilde c_{m+n},\qquad b_{mn}=c_m,\nonumber \\
c_m(q,t)&=& \frac{(1-q^m)(1-t^2q^{m-1})}{(1-tq^m)(1-tq^{m-1})}, \\
\tilde c_{m+n}(q,t)&=& \left( \frac{1-tq^{m+n}}{1-t^2q^{m+n}}\right)
\left( \frac{1-t^3q^{m+n-1}}{1-t^2q^{m+n-1}}\right) .\nonumber \end{eqnarray}
Note that
\be c_m(q,1)=c_m(q,q)=1,\qquad \tilde c_{m+n}(q,1)=\tilde c_{m+n}(q,q)=1. \ee

Now let us consider the case $t=q^k$, $k$ being an integer. In this case,
the explicit expression for the generating function of polynomials
$P_{n,0}^{(k)}\equiv P_{n,0}^{(q,t)}$ has the form
\begin{eqnarray}
G^{(k)}(u) &=& \prod _{j=0}^{k-1}\,F^1(q^ju),\nonumber \\
G^{(k)}(u) &=& \sum C_n^k\,P_{n,0}^{(k)}\,u^n, \\
C_n^k &=&\frac {[k]_n}{[1]_n},\qquad [x]_n=[x][x+1]\cdots [x+n-1],\quad [x]=
\frac{1-q^x}{1-x}. \nonumber \end{eqnarray}
From this we get a three-term recurrence relation being analogous to (4.7)
\be
[n+1]\,\tilde P_{n+1}^k=[k+n]\,z_1\,\tilde P_n^k-[2k+n-1]\,z_2\,
\tilde P_{n-1}^k +[3k+n-2]\,z_3\,\tilde P_{n-2}^k, \ee
which follows from the recurrence relation
\be \left( 1-z_1u+z_2u^2-z_3u^3\right) G^{(k)}(u) = \left( 1-z_1uq^k+z_2u^2
q^{2k}-z_3u^3q^{3k}\right) G(qu). \ee
From here, we obtain
\[
\tilde P_0^\kappa =1,\quad \tilde P_1^\kappa =[\kappa ]\,z_1,\quad
\tilde P_2^\kappa =\frac{[\kappa +1][\kappa ]}{[1][2]}\,z_1^2-
\frac{[2\kappa ]}{[2]}\,z_2,\,\ldots \,\quad
\tilde P _{m,0}=\frac{[1]_m}{[\kappa ]_m}\,P_{m,0},\]

Let us give also the formula for the generating function for the case of
arbitrary $t$:
\be G^{(q,t)}(u) = \prod _{j=0}^\infty \frac{(1-q^jtu)}{(1-q^ju)} =
\sum _{j=0}^\infty c_j(q,t)\,u^j. \ee

Using the results from the book [Ma 1995], it is not difficult to get
the formulae
\be
P_{m,0}^{(q,t)}\,P_{0,n}^{(q,t)}=\sum _{i=0}^{{\rm min}\,(m,n)}
\gamma ^i_{m,n}\,P_{m-i,n-i}^{(q,t)}\,, \ee
where
\be
\gamma ^i_{m,n}=\frac{(t;q)_i\,(q^m;q^{-1})_i\,(q^n;q^{-1})_i\,(q^{m+n-1-i}
t^3;q^{-1})_i}
{(q;q)_i\,(q^{m-1}t;q^{-1})_i\,(q^{n-1}t;q^{-1})_i\,(q^{m+n-i}t^2;q^{-1})_i};
\ee
and
\be
P_{m,0}^{(q,t)}\,P_{n,0}^{(q,t)}=\sum _{i=0}^{{\rm min}\,(m,n)}
\tilde \gamma _{m,n}^i\,P_{m+n-2i,i}^{(q,t)}\,, \ee
where
\be
\tilde \gamma ^i_{m,n} = \frac{(t;q)_i\,(q^m;q^{-1})_i\,(q^n;q^{-1})_i\,
(q^{m+n-i-1}t^2;q^{-1})_i}
{(q;q)_i(q^{m-1}t;q^{-1})_i(q^{n-1}t;q^{-1})_i(q^{m+n-i}t;q^{-1})_i}\,.\ee
Inverting formulae (B.10) and (B.11) as in the [PRZ 1998],  we come to

\noindent {\bf Theorem 1a.} {\em The Macdonald polynomials}
$P_{m,n}^{(q,t)}$ {\em of type} $A_2$ {\em are given by the
formula} \be P_{m,n}^{(q,t)}=\sum _{i=0}^{{\rm min} (m,n)}\beta
_{m,n}^i\,P_{m-i,0}^{(q,t)} \,P_{0,n-i}^{(q,t)}, \ee {\em where
constants have the form}
\begin{eqnarray}
\beta ^i_{m,n} &=& (-1)^i\,q^{i(i-1)/2}\,\frac{(t;q^{-1})}{(q;q)_i}\,
\frac{1-q^{m+n-2i}\,t^3}{1-q^{m+n-i}\,t^3}\nonumber \\
&&\,\times \,\frac{(q^m;q^{-1})_i\,(q^n;q^{-1})_i\,
(q^{m+n-1}t^3;q^{-1})_i}{(q^{m-1}t;q^{-1})_i\,(q^{n-1}t;q^{-1})_i\,
(q^{m+n-1}t^2;q^{-1})_i}\,. \nonumber \end{eqnarray}
In similar way, we get

\noindent{\bf Theorem  2a}. {\em The Macdonald polynomials} $P_{m,n}^{(q,t)}$
{\em of type} $A_2$ {\em are given by the formula}
\be
\tilde \gamma _{m+n,n}^n\,P_{m,n}^{(q,t)} = \sum _{i=0}^n
\tilde \beta _{m,n}^i\,P_{m+n+i,0}^{(q,t)}\,P_{n-i,0}^{(q,t)},\qquad m\geq n
\ee
{\em where constants $\tilde \gamma _{m,n}^i$ are given by formula} (B.13),
{\em and}
\be
\tilde \beta ^i_{m,n} = (-1)^i\,q^{i(i-1)/2}\,\frac{(t;q^{-1})_i}{(q;q)_i}\,
\frac{(1-q^{m+2i})}{(1-q^m)}\,\frac{(q^m;q)_i\,(q^n;q^{-1})_i\,
(q^{m+n}t;q)_i}{(q^{m+n+1};q)_i\,(q^{m+1}t;q)_i\,(q^{n-1}t;q^{-1})_i}.\ee

From Theorems 1a and 2a, many interesting identities may be obtained. Here
we give one of them:
\begin{eqnarray}
S_{n,l}^k&=& \sum _{i=0}^l(-1)^i\,q^{i(i-1)/2}\,\frac{[3k+n-2i]}{[3k+n-i]}
\left( \prod _{j=1}^i\frac{[k-j+1]}{[j]}\,\frac{[3k+n-j]}{[2k+n-j]}\right)
\nonumber \\
&&\times \left( \prod _{j=0}^{l-i-1}\frac{[k+j]}{[j+1]}\,
\frac{[3k+n-l-i-j-1]}{[2k+n-l-i-j]} \right) =0, \end{eqnarray}
where $[n]=(1-q^n)/(1-q)$.

The simplest version of this formula is
\be \sum _{i=0}^l (-1)^i\,q^{i(i-1)/2}\frac{[k+l-i-1]}{[i]!\,[l-i]!\,[k-i]!}
=0. \ee

We give below the {\bf list of polynomials} $P_{m,n}^{(q,t)}$ at $m+n\leq 4$
\begin{eqnarray*}
P_{0,0}^{(q,t)}&=& 1,\qquad P_{1,0}^{(q,t)}=z_1,\qquad P_{0,1}^{(q,t)}=z_2,\\
P_{2,0}^{(q,t)} &=& z_1^2-\frac{(1-q)(1+t)}{(1-qt)}\,z_2,\\
P_{1,1}^{(q,t)} &=& z_1z_2-\frac{(1-q)(1+t+t^2)}{(1-qt^2)}\,z_3,\\
P_{3,0}^{(q,t)} &=& z_1^3-\frac{(1-q)(2+q+t+2qt)}{1-q^2t}\,z_1z_2+
\frac{(1-q)^2(1+q)(1+t+t^2)}{(1-qt)(1-q^2t)}\,z_3, \end{eqnarray*}
\begin{eqnarray*}
P_{2,1}^{(q,t)} &=& z_1^2z_2-\frac{(1-q)(1+t)}{1-qt}\,z_2^2+\frac{(1-q^2)
(1-qt^3)}{(1-qt)^2(1+qt)}\,z_1z_3,\\
P_{4,0}^{(q,t)} &=&
z_1^4-\frac{(1-q)(3+2q+q^2+t+2qt+3q^2t)}{1-q^3t}\,z_1^2z_2\\
&&+\,\frac{(1-q)^2(1+q+q^2)(1+t)(1+qt)}{(1-q^2t)(1-q^3t)}\,z_2^2\\
&&+\,\frac{(1-q)^2(1+q)(2+q+q^2+t+2qt+q^2t+t^2+qt^2+2q^2t^2)}{(1-q^2t)
(1-q^3t)}\,z_1z_3,\end{eqnarray*}
%or
%\be \cdots \frac{(1-q)^2(1+q)((q+t)(1+q+t+qt)+2(1+q^2t^2))}{\cdots }
\begin{eqnarray*}
P_{3,1}^{(q,t)} &=& z_1^3z_2-\frac{(1-q)(2+q+t+2qt)}{1-q^2t}\,z_1z_2^2 -
\frac{(1-q^3)(1-q^2t^3)}{(1-q^2t)(1-q^3t^2)}\,z_1^2z_3\\
&&+\,\frac{(1-q)^2}{(1-q^2t)(1-q^3t^2)} \\
&&\times \,\left( 2+2q+q^2+2t+4qt+3q^2t+q^3t+t^2+\right. \\
&&\,\,\,\left.3qt^2+4q^2t^2+2q^3t^2+qt^3+2q^2t^3+2q^3t^3\right) z_2z_3;
\end{eqnarray*}
% P_{3,1}=\cdots \times ((2+2q+q^2)(1+(1+q)t)+t^2(1+2q+2q^2)(1+q+qt))
\begin{eqnarray*}
P_{2,2}^{(q,t)}&=& z_1^2z_2^2-\frac{(1-q)(1+t)}{1-qt}\,(z_2^3+z_1^3z_3)\\
&&-\,\frac{(1-q)(3q+q^2-3t+2q^2t+q^3t-t^2-2qt^2+3q^3t^2-qt^3-3q^2t^3)}
{(1-qt)(1-q^3t^2)}\,z_1z_2z_3\\
%P_{22}=\cdots \times (1-q)2(1+q)(1+t+t^2)(q-t)(1+qt+q^2t^2)
&&+\,\frac{(1-q)^2(1+q)(1+t+t^2)(q-t+q^2t-qt^2+q^3t^2-q^2t^3)}
{(1-qt)(1-q^2t^2)(1-q^3t^2)}\,z_3^2. \end{eqnarray*}
%%%%%
%% \section*{Acknowledgment(s)}
%% text of Acknowledgement

%% \begin{thebibliography}{**}
%% \bibitem{}
%% ...
%% \end{thebibliography}

%%%%
\section*{Acknowledgment}

The paper was complete during my stay at Max-Planck-Institut
f\"ur Mathematik Bonn.
I am grateful the members of the Institute for hospitality.

\end{document}